\documentclass[12pt]{article}

\usepackage{graphicx}
\usepackage{amsmath}
\usepackage[letterpaper,margin=1in]{geometry}
\renewenvironment{abstract}
	{\quotation}
	{\endquotation}

\date{}

\makeatletter
\renewcommand{\fnum@figure}{\textbf{Figure \thefigure}}
\renewcommand{\fnum@table}{\textbf{Table \thetable}}
\makeatother

\usepackage{scicite}

\usepackage{url}

\def\scititle{
	On the photon energy conservation in stimulated emission. Experiment
}

\title{\bfseries \boldmath \scititle}

\author{
	Pavel L. Chapovsky,\and
	\small Institute of Automation and Electrometry SB RAS, Novosibirsk \& 630090, Russia.\and
	\small Institute of Laser Physics SB RAS, Novosibirsk \& 630090, Russia.\and
	\small International Tomography Center SB RAS, Novosibirsk \& 630090, Russia. \and
	\small Email: chapovsky@iae.nsk.su
}

\begin{document}

\maketitle

\begin{abstract} \bfseries \boldmath
Quantum electrodynamics predicts identity of incident and emitted photons in stimulated emission. This fundamental law is important to test experimentally. In this work stimulated emission in GaAs semiconductor amplifier was investigated and positive frequency shift
of the amplified beam was detected. In relative units this frequency shift was found equal $\Delta\nu/\nu = (+1.7 \pm 0.2)\cdot10^{-18}$.
This indicates violation of the photon energy conservation in stimulated emission.
\end{abstract}

\section*{Introduction}

Foundations of quantum theory of radiation interaction with matter were created by Einstein in 1916 \cite{Einstein1916VDPG,Einstein1916MPG}.
In his model quantized electromagnetic field induces in molecules processes of three types: absorption,
spontaneous emission and stimulated emission. Absorption is the easiest to understand process.
It has also a clear classical analogue.

Spontaneous emission is a quantum process. In classical analogue of this process
an excited oscillator emits "spherical" wave. In Einstein's theory molecule emits a photon in particular direction.
"Spherical" symmetry of emitted photons appears on average only. In modern terms Einstein's spontaneous emission
describes a molecule interaction with quantized vacuum.

The third process is stimulated emission of photons. This is a most complicated process which has no classical
analog. Its closest classical counterpart describes radiation emission by an oscillator excited by light.
In classical theory one needs to spend radiation energy for the oscillator excitation. In quantum theory it is not.
Main attention in works \cite{Einstein1916VDPG,Einstein1916MPG} was devoted to the matter
modification by radiation and to the derivation of Planck distribution for radiation spectral density. One can conclude also that
in these papers Einstein stated that the emitted photon in stimulated emission is identical to the incident photon \cite{Masters12OPN}.

This conclusion was confirmed by quantum electrodynamics developed by Dirac \cite{Dirac27PRSLA}.
Quantum electrodynamics describes the stimulated emission as emission a photon at the same state which the incident photon has.
There are some arguments that the two photons in stimulated emission may be not perfectly identical.
Einstein deduced his theory by condition that the light-matter interaction conserves Boltzmann distribution
for the molecular level populations and Maxwell distribution for the molecular momentums.
Note that the Boltzmann and Maxwell distributions are classical ones and not
universal. Quantum mechanics modifies these distributions. For example, such modification appears in the case of Bose-Einstein condensation. Quantum electrodynamics states that one cannot produce a perfect copy of quantum system due to the
"no-cloning theorem" \cite{Wootters82N}. Contemporary works on quantum informatics investigate to what extent the photon cloning by stimulated emission can be perfect \cite{Lamas-Linares02S}. In this work we address the problem of equality of photons' energies in stimulated emission.

\section*{Experiment}

\begin{figure}
	\centering \includegraphics[width=0.8\textwidth]{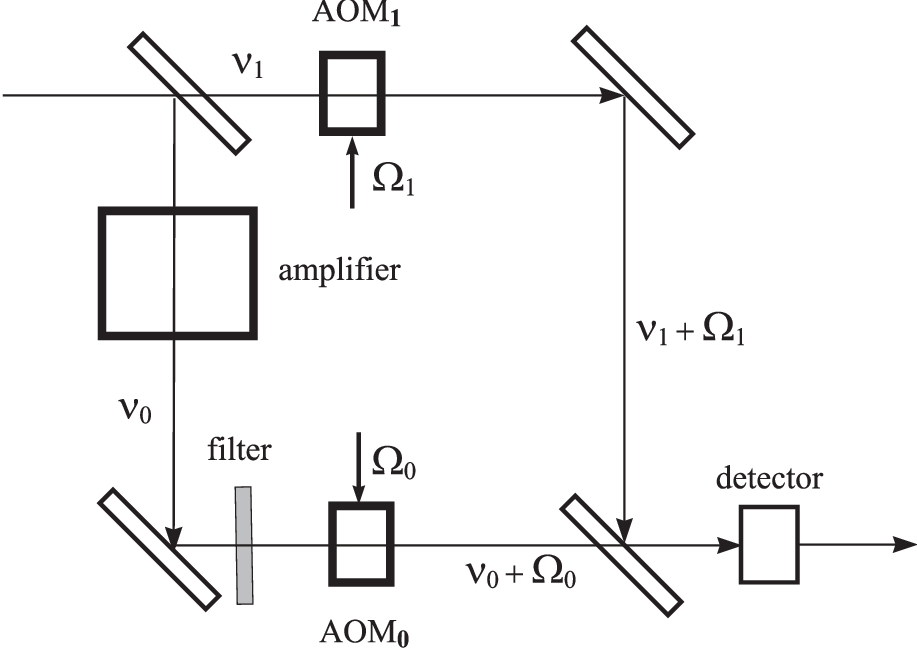}
	\caption{\textbf{Principle of the experimental setup.} Experiment is based on the use of Mach-Zander interferometer.
AOM is an acousto-optical modulator. Amplifier is GaAs semiconductor tapered amplifier (Eagleyard Photonics, Germany).}
	\label{fig1}
\end{figure}

Properties of stimulated emission were studied in this work using GaAs semiconductor amplifier.
The tapered amplifier (Eagleyard Photonics, Germany) was installed in one arm of Mach-Zehnder interferometer (Fig.~\ref{fig1}). Another interferometer arm contained unamplified laser beam. In order to increase frequency sensitivity of optical beats at the interferometer output the acousto-optical modulators (AOMs) were installed into the two interferometer arms. We fed these AOMs by RF signals having frequencies near 76~MHz but with small frequency shift between the two.

Setup in Fig.~\ref{fig1} allowed us efficient suppression of the amplifier's spontaneous radiation at the interferometer output. Mach-Zehnder interferometer is especially suitable for the measurements of optical beats which are related to
the energy difference between amplified and unamplified photons. The experiment was designed to exclude stationary phase shift and amplitude imbalance between interferometer arms.

Let us denote the light frequencies at the two interferometer arms as $\nu_0$ and $\nu_1$ (Fig.\ref{fig1}). The AOMs shifted
the light frequencies by $\Omega_0$ and $\Omega_1$. Consequently, the optical beats frequency at the interferometer output was
$\delta'_{opt} = \nu_0 + \Omega_0 - (\nu_1 + \Omega_1) = \nu_0-\nu_1 - (\Omega_1-\Omega_0)$.
The goal of the experiment was to measure the frequency shift of the amplified light beam $\Delta\nu=\nu_0-\nu_1$. Our experimental method allowed us to measure only the absolute frequency differences $\delta_{RF}=|\Omega_1-\Omega_0|$ and $\delta_{opt}=|\delta'_{opt}|$. In our experimental conditions $\delta_{RF}\gg |\nu_0-\nu_1|$. In this limit, the optical frequency shift $\Delta\nu$ is determined by the expression
\begin{equation}\label{Dn}
\Delta\nu = \eta(\Omega_1-\Omega_0)\cdot(\delta_{RF}-\delta_{opt}).
\end{equation}
Here, the function $\eta(x)=1$ at~$ x > 0$ and $\eta(x)=-1$ at~$ x < 0$. An accuracy of the measurements of $\delta_{opt}$ was hardware dependent. The highest accuracy was obtained at our smallest possible RF frequency shift $\delta_{RF}\simeq0.11$~Hz.

The setup presented in Fig.~\ref{fig1} is simplified one. Actual experimental setup was more complicated.
It contained additional elements: focussing and collimating optics for the tapered amplifier,
two Faraday isolators which prevented back light reflection to the laser and to the tapered amplifier.
Consequently, it was not possible to simply remove the tapered amplifier from the setup in order to measure
the beats between two unamplified beams. The beats of unamplified beams were measured by
modified Mach-Zehnder interferometer. We took a special care in order to reduce the technical noises caused by
the optical table vibration and air flows.

As a light source we used continuous semiconductor laser: model DL100, Toptica Photonics (Germany).
The laser wavelength was 780~nm, radiation power 50~mW and radiation linewidth 0.5~MHz. Laser radiation frequency was stabilized
on the center of hyperfine transition F$_g = 3\rightarrow$ F$_e = 4$ of D$_2$ line of Rubidium isotope $^{85}$Rb by DAVLL stabilization method \cite{Corwin98AO}. The frequency stabilization accuracy was 1~MHz.
Central wavelength of the tapered amplifier was 780~nm, amplification width (FWHM) 20~nm. Maximum power of the amplifier output was 0.5~W.

\begin{figure}
	\centering \includegraphics[width=0.8\textwidth]{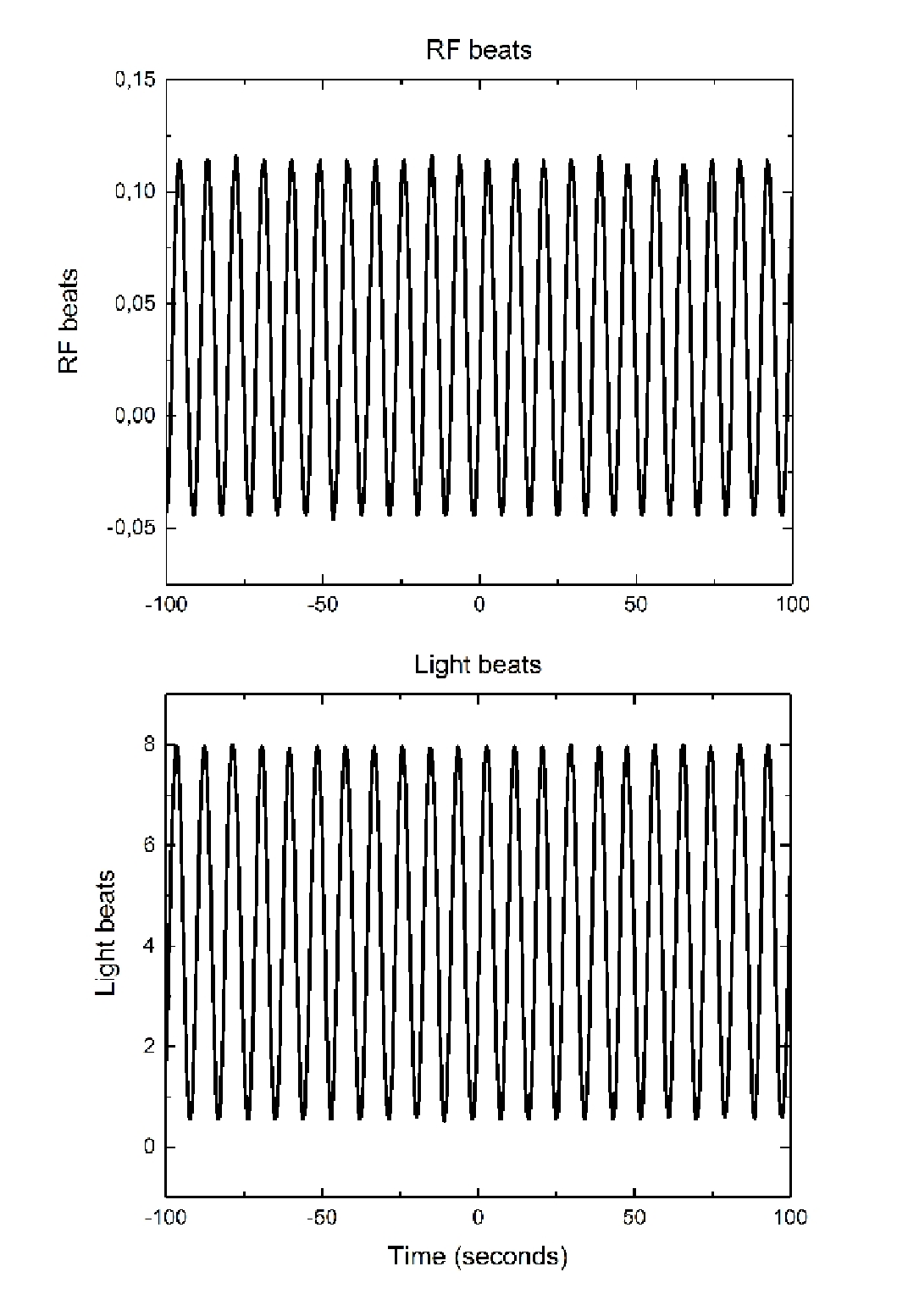}
	\caption{\textbf{RF and optical beats.}
		\textbf{Upper panel}: Fragment of the RF beats at $\delta_{RF}\simeq0.11$~Hz. \textbf{Low panel}: Fragment of the optical beats
at $\delta_{RF}\simeq0.11$~Hz.}
	\label{fig2}
\end{figure}

We used Direct Digital Synthesis (DDS) technology for generation RF signals. It was done by AD9959 evaluation board
and software from Analog Devices (USA). This device generated four analog RF signals having
extremely high frequency stability relative to each other. This stability appears because the DDS converts
digital RF signals to analog RF signals using common reference clock (having 480~MHz in our case). Relative stability of the
generated RF signals was determined by measuring their beats. The beats were recorded at 2500 points using
digital storage oscilloscope TDS2024 which served as a local frequency standard in our setup. The record duration was 500 seconds for the frequency shift of the two AOMs equal $\delta_{RF}\simeq0.11$~Hz. An example of RF beats is presented in Fig.~\ref{fig2} (upper panel).

A dedicated software was used to process the beats numerically. We used the fitting function
\begin{equation}\label{y}
y(t) = a\cdot\sin(2\pi\delta\cdot t + b) + c,
\end{equation}
where $t$ is the time in seconds and $a,\, \delta,\, b,\, c$ are the fitting parameters. Achieved statistical uncertainty of the RF beats was $0.5~\mu$Hz (95\%
confidence interval here and in all similar places in this work).
Most probably, the actual relative stability of two RF signals was even better than $0.5~\mu$Hz. The statistical uncertainty was hardware imposed because of the limited number of recorded points, record time duration and RF beats detection accuracy. It was important that the achieve frequency accuracy of the RF beats was sufficiently high for the quantitative analysis of the optical beats (see, below).
Example of the optical beats at $\delta_{RF}\simeq0.11$~Hz is presented in Fig.~\ref{fig2} (low panel). The optical beats were recorded and processed similar to the RF beats. Each record of the optical beats was accompanied with simultaneous record of the RF beats in order to determine the optical frequency shift by the Equation~(\ref{Dn}).

\section*{Experimental Results}

Highest resolution of the optical beats frequency $\delta_{opt}$ was achieved in our setup at smallest available frequency difference of the RF signals generated by DDS, $\delta_{RF}\simeq0.11$~Hz. The actual magnitude of this $\delta_{RF}$ was carefully measured and
was found in a single record of the RF beats equal to
\begin{equation}\label{dRF}
 \Omega_1 - \Omega_0 = 0.1117577~\text{Hz}\pm0.5~\mu\text{Hz}.
\end{equation}
In order to measure the total error of $\delta_{RF}$ that combines both statistical and systematic errors the measurements of $\delta_{RF}$ were repeated for a few days (27 RF beats records). As the result, the error of $\delta_{RF}$ was found equal to $\simeq1~\mu$Hz.

\begin{figure}
	\centering \includegraphics[width=1\textwidth]{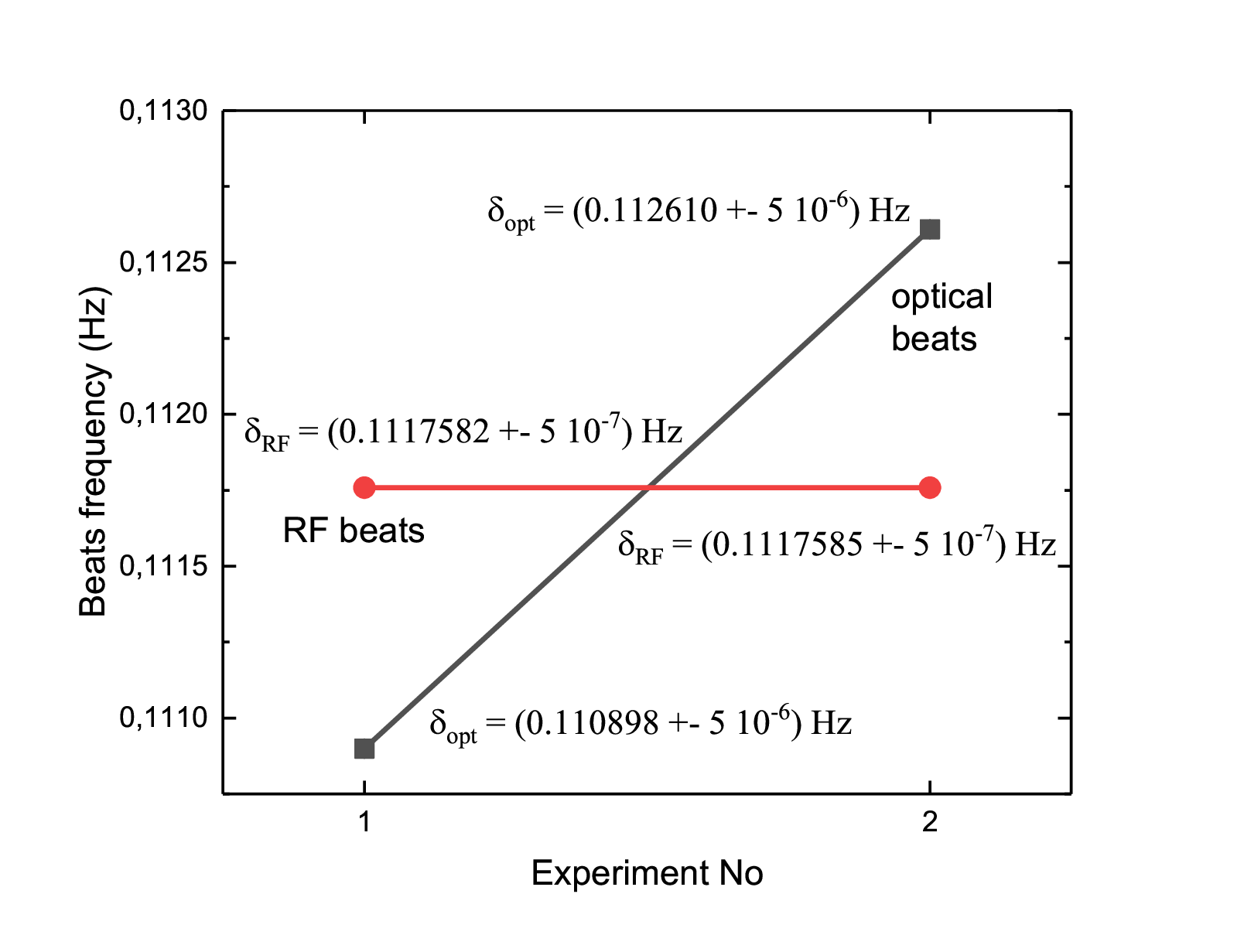}
	\caption{\textbf{Frequency measurements of RF and optical beats.}
     Filled circles refer to the RF beats. Filled squares refer to the optical beats.
     The lines are added to guide the eye.}
	\label{fig3}
\end{figure}

Let us turn now to the measurements of the optical beats recorded at $\delta_{RF}\simeq0.11~$Hz.
The measurements of the frequency difference of two unamplified optical beams resulted in
\begin{equation}\label{LL}
\Delta\nu_{un}\equiv (\nu_0-\nu_1)_{un} = (-56 \pm 73)~\mu\text{Hz},
\end{equation}
and thus did not show a measurable frequency shift. Here, the statistical plus systematic error is indicated.
Contrary to this, the amplified beam did show the frequency shift.
First we discuss the experiment that tested the influence of the sign of $\Omega_1-\Omega_0$ on the amplified beam frequency shift, $\Delta\nu$.  Parameters of the two experiments different by the sign of $\Omega_1-\Omega_0$ are given in Table~\ref{Table1} and in Fig.~\ref{fig3}. Here, statistical errors of particular measurements are given. One can see that we obtained very close frequency shifts of the amplified beam in these two
experiments. As one would expect the measured frequency shift of the amplified beam $\Delta\nu$ did not depend on the sign of AOM frequency differences, $\Omega_1-\Omega_0$.

Collection of our measurements at $\delta_{RF}\simeq0.11$~Hz (27 records made during a few days) resulted in the frequency difference between amplified and unamplified beams equal
\begin{equation}\label{Dnu}
  \Delta\nu\equiv \nu_0 - \nu_1 = (+640 \pm 80)~\mu\text{Hz}.
\end{equation}
Thus, the amplified beam had positive systematic frequency shift. To the best of our knowledge this is the first
observation of the frequency shift of the amplified beam in stimulated emission.

\section*{Conclusions}
We have registered low frequency ($\simeq0.11$~Hz) optical beats in Mach-Zehnder interferometer.
In view of relatively large spectral width of the laser radiation ($\simeq0.5$~MHz), this low frequency optical beats can be interpreted
as a photon interference with itself.

We did registered a systematic positive shift of the frequency of amplified optical beam relative to unamplified optical beam equal to
$\Delta\nu = (+640 \pm 80)~\mu$Hz  and in relative units
\begin{equation}\label{Dnn}
\Delta\nu/\nu = (+1.7 \pm 0.2)\cdot10^{-18}.
\end{equation}
This frequency shift indicates violation of the photon energy conservation in stimulated emission.

\section*{Acknowledgments}
The author is indebted to participants of the Russian conference on Physics of Ultracold
Atoms (Novosibirsk, December 2024) where this work was presented.

All data are available in the manuscript.

\clearpage
\begin{table} % Do NOT use \begin{table*}
	\centering
	% Captions go above tables
	\caption{\textbf{Influence of the sign of $\Omega_1-\Omega_0$.}
		Here $\Omega_1$ and $\Omega_0$ are the frequencies of RF signals fed to AOMs.
        $\delta_{opt}$ is the frequency of optical beats.
        $\Delta\nu$ is the frequency shift between amplified and unamplified optical beams.}
	\label{Table1} % give each table a logical label name
	
	\begin{tabular}{lccc} % four columns, alignment for each
		\\
		\hline
		Exp. & $\Omega_1-\Omega_0$          & $\delta_{opt}$            & $\Delta\nu$                  \\
		No   &     (Hz)                     &   (Hz)                          & (Hz)                               \\
		\hline
		1    & $0.1117582\pm5\cdot10^{-7}$  & $0.110898\pm5\cdot10^{-6}$      & $8.60\cdot10^{-4}\pm5\cdot10^{-6}$ \\
		2    & $-0.1117585\pm5\cdot10^{-7}$ & $0.112610\pm5\cdot10^{-6}$      & $8.52\cdot10^{-4}\pm5\cdot10^{-6}$ \\
		\hline
	\end{tabular}
\end{table}

\clearpage

\clearpage

\clearpage

\end{document}